\documentclass[twocolumn,A4]{article}
\usepackage[dvips]{graphics}
\begin{document}
\onecolumn
\begin{center}
{\bf{\Large Enhancement of persistent current in mesoscopic rings $\&$
cylinders : shortest $\&$ next possible shortest higher
order hopping}} \\
~\\
Santanu K. Maiti$^*$, J. Chowdhury and S. N. Karmakar \\
~\\
{\em Saha Institute of Nuclear Physics, 1/AF, Bidhannagar, 
Kolkata 700 064, India} \\
~\\
~\\
~\\
{\bf Abstract}
\end{center}

 We present a detailed study of persistent current and low-field magnetic
susceptibility in single isolated normal metal mesoscopic rings and cylinders 
in the tight-binding model with higher order hopping integral in the 
Hamiltonian. Our exact calculations show that order of magnitude 
enhancement of persistent current takes place even in presence of disorder 
if we include higher order hopping integral in the Hamiltonian. 
In strictly one-channel mesoscopic rings the sign of the low-field currents 
can be predicted exactly even in presence of impurity. We observe that
perfect rings with both odd and even number of electrons support only
diamagnetic currents. On the other hand in the disordered rings, 
irrespective of realization of the disordered configurations of the ring, 
we always get diamagnetic currents with odd number of electrons and 
paramagnetic currents with even number of electrons. In mesoscopic cylinders
the sign of the low-field currents can't be predicted exactly since it
strongly depends on the total number of electrons, $N_e$, and also on the
disordered configurations of the system. From the variation of persistent
current amplitude with system size for constant electron density, we 
conclude that the enhancement of persistent current due to additional 
higher order hopping integrals are visible only in the mesoscopic regime.
\vskip 2cm
\begin{flushleft}
\end{flushleft}
\vskip 1.5in
\noindent
{\bf ~$^*$Corresponding Author}:

Email address: santanu.maiti@saha.ac.in 
\newpage
\twocolumn
\section{Introduction}
 Since $1960$ the study of magnetic response in normal metal mesoscopic
loops provides many exotic new results as a consequence of phase coherence 
of the electrons in these small scale systems. B\"{u}ttiker, Imry 
and Landauer~\cite{butt} have shown following the works of Byers and 
Yang~\cite{byers} that a small isolated normal metal ring threaded by 
slowly varying magnetic flux $\phi$ carries an equilibrium current and
it never decays, even in presence of impurity in the system. This current
varies periodically with $\phi$ showing $\phi_0$ flux-quantum periodicity. 
Later experimental results have verified the existence of persistent current 
in such small rings. At very low temperatures, the inelastic scattering 
length is much larger than the ring size, $L$, and accordingly, the electron 
transport is completely phase coherent throughout the ring. Again in these 
small systems with finite size the energy levels are discrete. The large phase 
coherence length, $L_(\phi)$, and the discreteness of the energy levels play 
an important role in the existence of persistent current in these normal
metal loops. There are lots of theoretical studies~\cite{cheu1,cheu2,mont,
alts,von,schm,ambe,abra,bouz,giam,san1} on persistent current in 
normal metal rings, but till now we are unable to explain many features 
of these currents that are observed experimentally. The experimental 
results on single isolated rings are significantly different from those for 
the ensemble of single isolated rings. The measured average currents are 
comparable to the sample-specific typical currents $<I^2>^{1/2}$ predicted 
for a single ring, but are one or two orders of magnitude larger than the 
ensemble averaged persistent currents expected from free electron theory. The
theoretical calculations including electron-electron interaction yield
average persistent current within an order of magnitude of the experimental 
value, but cannot explain the diamagnetic sign of the currents.  
Levy {\em et al.}~\cite{levy} have measured diamagnetic response of the 
currents at very low fields in an experiment on $10^7$ isolated mesoscopic 
Cu rings. On the other hand, Chandrasekhar {\em et al.}~\cite{chand} have 
determined $\phi_0$ periodic currents in Ag rings with paramagnetic 
response at low fields. On the theoretical side, 
Cheung {\em et al.}~\cite{cheu2} predicted that the direction of persistent 
current is random depending on the total number of electrons, $N_e$, in the 
system and the specific realization of the random potentials. Both diamagnetic 
and paramagnetic responses have been observed theoretically in mesoscopic 
Hubbard ring by Yu and Fowler~\cite{yu}. They have shown that the rings with 
odd $N_e$ exhibit paramagnetic response while those with even $N_e$ give
diamagnetic response in the limit $\phi \rightarrow 0$. In a recent experiment 
Jariwala {\em et al.}~\cite{jari} obtained diamagnetic persistent currents 
with  both $\phi_0$ and $\phi_0/2$ flux-quantum periodicities in an array of 
$30$-diffusive mesoscopic gold rings. The diamagnetic sign of the currents 
in the vicinity of zero magnetic field were also found in an 
experiment~\cite{deb} on $10^5$ disconnected Ag ring. The sign is a priori 
not consistent with the theoretical predictions for the average of persistent 
current. Thus we see that theory and experiment still do not agree very well. 

\begin{figure}[ht]
{\centering \resizebox*{5.75cm}{3.3cm}{\includegraphics{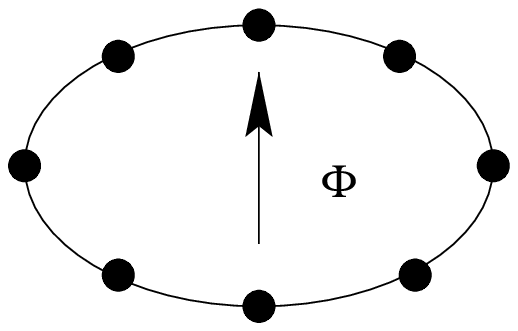}}\par}
\caption{One dimensional normal metal ring threaded by a magnetic flux $\phi$.
The filled circles denote the positions of the lattice site.}
\label{ring}
\end{figure}

In this article we shall describe magnetic response of one-dimensional
normal metal mesoscopic rings and cylinders within the one-electron picture 
using tight-binding Hamiltonian. All most all the existing theories are 
basically based on the framework of nearest-neighbor tight-binding 
Hamiltonian with either diagonal 
disorder or off-diagonal disorder. But here we consider an additional 
higher order hopping integral with the nearest-neighbor hopping (NNH) integral 
in the Hamiltonian and try to explain the dependences of persistent 
currents on the number of electrons $N_e$, disorder strengths $W$. 
We can consider higher order hopping integrals in the Hamiltonian on the
basis that the overlap of the atomic orbitals between various neighboring 
sites are usually non-vanishing and the higher order hopping integrals 
become quite important. In this article we take only one higher order 
hopping integral, in addition to the NNH integral, which gives the hopping
of an electron in the {\em next shortest path} between two sites.
In case of strictly one-dimensional rings, i.e., rings with only one channel 
the next possible {\em shortest path} is equal to the twice of the lattice 
spacing (see Fig.~\ref{ring}), while, in cylinders it would be the diagonal 
distance (shown by the arrows in Fig.~\ref{cylinder}) of each small 
rectangular loop (see Fig.~\ref{cylinder}).
 
 This paper is organized as follows. In section II, we study the variation 
of persistent current as a function of magnetic flux $\phi$ in strictly
one-dimensional mesoscopic rings. Here we describe the 
dependences of persistent currents on electron numbers $N_e$, disorder 
strengths $W$ and also on higher order hopping integral. 
Section III describes the behavior of persistent currents in mesoscopic
cylinders and the effects of diagonal hopping integral on current in 
presence of impurity. The sign of the low-field currents in these mesoscopic
rings and cylinders is clearly investigated in section IV. In section V, we
compute the variation of current amplitudes with system size $N$ both for
one-channel mesoscopic rings and cylinders. Lastly, our conclusions are 
given in section VI.
\section{One-dimensional Mesoscopic Ring}
 The Hamiltonian for a $N$-site ring in the tight-binding model can be 
written as, 
\begin{equation}
H=\sum_{i}\epsilon_i c_i^{\dagger} c_i + \sum_{i\ne j}v_{ij}
\left[e^{i\theta_{ij}} c_i^{\dagger}c_j + h.c. \right]  
\label{hamil}
\end{equation}
where $\epsilon_i$'s are the site potential energies and the phase factors 
are $\theta_{ij}={2\pi\phi\left(|i-j|\right)}/N$. We take the hopping 
integral between any two sites $i$ and $j$, here in our present model $j$ 
has the values $i\pm 1$ and $i \pm 2$ only, in the form 
$v_{ij}=v\exp\left[\alpha(1-|i-j|)\right]$, where $v$ is the hopping strength 
between any two nearest-neighbor sites. In this work we use the units 
$c=e=h=1$. As we are considering only non-magnetic impurities, the spin of 
the electrons will not produce any qualitative change in the behavior of 
persistent current and low-field magnetic susceptibility, and so we neglect 
the spin of the electrons throughout this work. 

 At zero temperature, the persistent current is given by
\begin{equation}
I(\phi)=-\frac{\partial E(\phi)}{\partial \phi}
\end{equation}
where $E(\phi)$ is the ground state energy of the system. For a perfect
ring we can calculate ground state energy analytically, while in 
a disordered ring we do exact numerical diagonalization to evaluate the 
ground state energy. Gauge invariance~\cite{byers} implies that $I(\phi)$ 
is a periodic function of $\phi$ with period $\phi_0=ch/e=1$.

 In this section we investigate the behavior of current-flux characteristics 
both for the ordered and disordered rings described by the Hamiltonians with 
only NNH integral and the rings described by the Hamiltonians with NNH 
integral in addition to the second neighbor hopping (SNH) integral. 
For an ordered ring we put $\epsilon_i=0$ for all $i$ in the above Hamiltonian 
given by Eq.~(\ref{hamil}), and the energy of the $n$th single-particle state 
can be expressed as
\begin{equation}
E_n(\phi)=\sum_{p=1}^{p_0} 2~v\exp\left[\alpha (1-p)\right]
\cos\left[\frac{2\pi p}{N} \left(n+\phi\right)\right]
\label{engequ1}
\end{equation}
and the current carried by this eigenstate is
\begin{eqnarray}
I_n(\phi) &=& \left(\frac{4\pi v}{N}\right)\sum_{p=1}^{p_0} 
p~\exp\left[\alpha (1-p)\right] \times \nonumber \\
& & \sin\left[\frac{2\pi p}{N}\left(n+\phi\right)\right]
\label{currequ1}
\end{eqnarray}
where $p$ is an integer. We take $p_0=1$ and $2$ respectively for the 
rings with NNH and SNH integrals.  
\begin{figure}[ht]
{\centering \resizebox*{8.65cm}{9cm}{\includegraphics{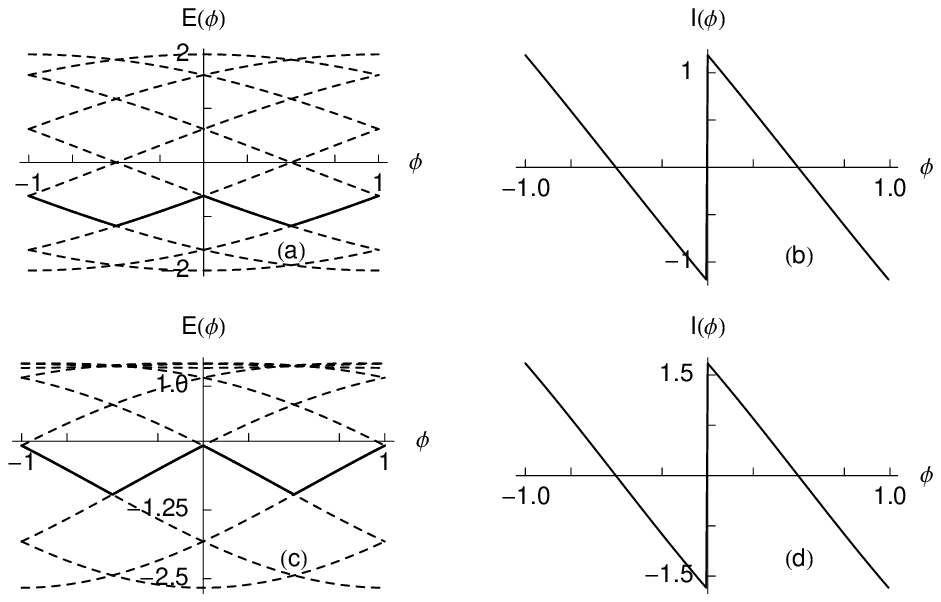}}\par}
\caption{Energy spectra and persistent currents of $10$-site perfect rings 
with four electrons ($N_e=4$), where a) and b) correspond to NNH
model while c) and d) correspond to SNH ($\alpha=1.1$) model.} 
\label{fermiorder}
\end{figure}  

 At zero temperature, we can write the total persistent current in the
following form
\begin{equation}
I(\phi)=\sum_n I_n(\phi)
\end{equation} 
where $n$ is an integer and restricted in the range $-\lfloor N_e/2\rfloor 
\leq n < \lfloor N_e/2 \rfloor$ ($\lfloor z\rfloor$ denotes the integer 
part of $z$), where $N_e$ denotes the number of electrons. 

 For large values of $\alpha$, the systems described by the SNH 
integral eventually reduce to the systems with only NNH integral. 
As we decrease the value of $\alpha$, 
\begin{figure}[ht]
{\centering \resizebox*{8.5cm}{9cm}{\includegraphics{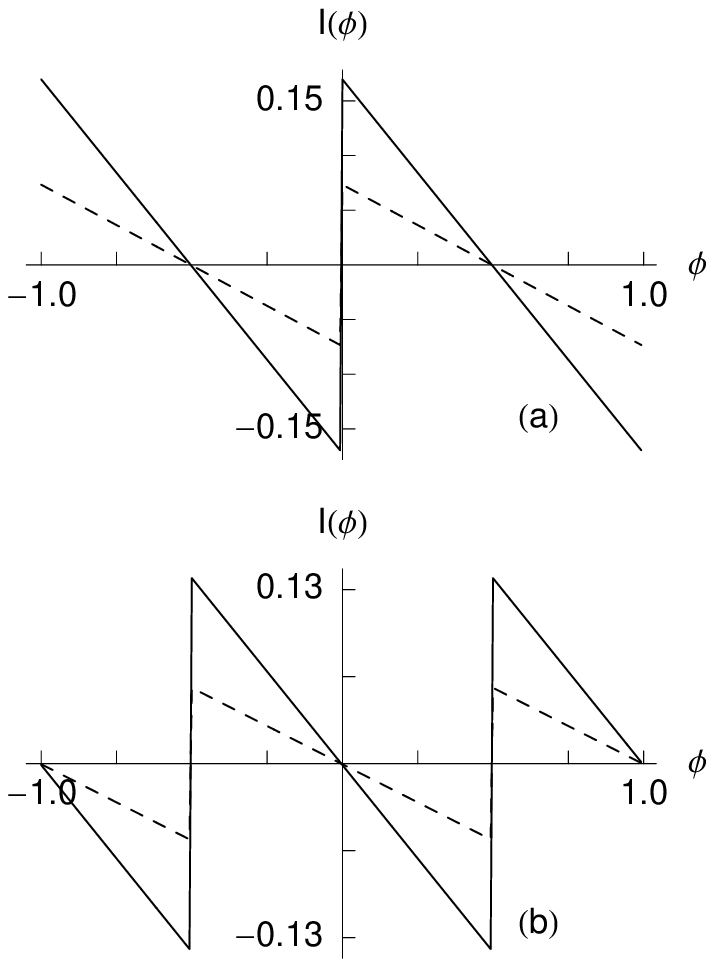}}\par}
\caption{Persistent current as a function of $\phi$ for ordered rings 
with $N=100$, $\alpha=0.9$, and, a) $N_e=20$ and b) $N_e=15$. The
dotted and solid lines are respectively for the rings with NNH and 
SNH integrals.}
\label{current1}
\end{figure}  
contributions from the SNH integral become much more appreciable, and, the 
energy spectrum and persistent currents get modified, and these modifications 
give some new results both in absence and presence of disorder in the systems. 

 To reveal this fact, we first present in Fig.~\ref{fermiorder} the energy
spectra and persistent currents of $10$-site perfect rings with four 
electrons. The energy spectra for the NNH and SNH models are respectively
shown in Fig.~\ref{fermiorder}(a) and Fig.~\ref{fermiorder}(c), and the 
solid curves give the variation of Fermi level at $T=0$ with flux $\phi$.
We see that the SNH integral lowers the energy levels and most importantly
below Fermi level the slopes of the $E(\phi)$ versus $\phi$ curves 
increases. As a result persistent current increases in the presence of SNH
integrals and this enhancement of persistent current is clearly visible 
from Fig.~\ref{fermiorder}(b) and Fig.~\ref{fermiorder}(d). 
In Fig.~\ref{fermiorder} we have considered $10$-site rings only for the 
sake of illustration and the results for the larger rings are presented 
in Fig.~\ref{current1}. 

 In Fig.~\ref{current1} we plot $I(\phi)$ versus $\phi$ curves for some 
perfect rings with $N=100$ and $\alpha=0.9$. The dotted and solid lines 
respectively gives the variation of current as a function of magnetic flux 
$\phi$ for the systems with NNH and SNH integrals. The enhancement of
current amplitudes due to the addition of SNH integral is clearly observed
from Fig.~\ref{current1}(a) and Fig.~\ref{current1}(b) if we compare the
results plotted by the dotted and solid curves. Fig.~\ref{current1}(a)
shows that current has sharp transitions at $\phi=0$ or $\pm n\phi_0$,
while, in Fig.~\ref{current1}(b) current shows the transitions at 
$\phi=\pm n\phi_0/2$. These transitions are due to the degeneracy of the 
energy eigenstates at these respective fields. From Fig.~\ref{current1} 
we see that for all the above models persistent currents are always periodic 
in $\phi$ with $\phi_0$ flux periodicity.

\begin{figure}[ht]
{\centering \resizebox*{8.75cm}{9.5cm}{\includegraphics{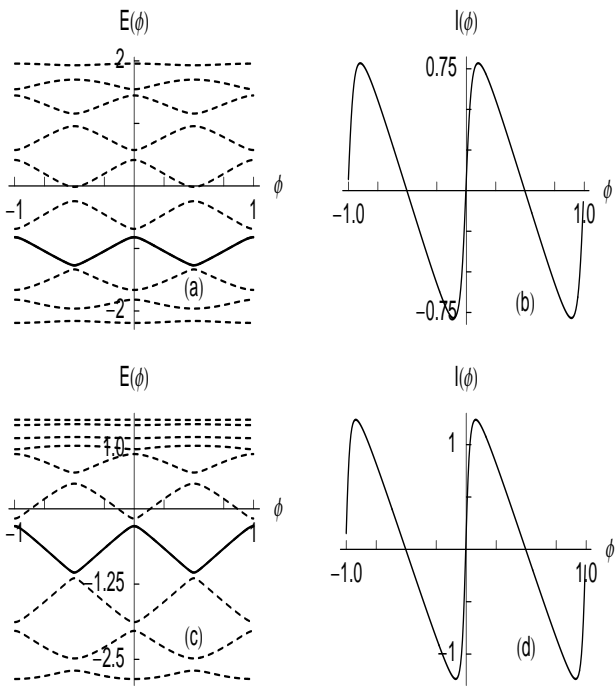}}\par}
\caption{Energy spectra and persistent currents of $10$-site disordered 
($W=1$) rings with four electrons ($N_e=4$), where a) and b) correspond to
NNH model while c) and d) correspond to SNH ($\alpha=1.1$) model.} 
\label{fermidisorder}
\end{figure}

 To understand the role of higher order hopping integral on persistent 
currents in disordered rings, we first study the energy spectra and currents
in small rings, and the results for $10$-site rings with $N_e=4$ are shown
in Fig.~\ref{fermidisorder}. We describe the system by Hamiltonian
Eq.~(\ref{hamil}) with site energies $\epsilon_i$'s chosen randomly between
$-W/2$ to $W/2$, where $W$ being the strength of disorder. In 
Fig.~\ref{fermidisorder}(a) and Fig.~\ref{fermidisorder}(c) we present the
energy spectra respectively for the NNH and SNH models where solid curves
give the location of the Fermi level. As in the ordered situations, the SNH
integral lowers the energy levels and below Fermi level the slopes of the 
$E(\phi)$ versus $\phi$ curves are much more than those for the NNH model.
Thus even in the presence of disorder, we have enhancement of persistent 
current due to SNH integral. 

\begin{figure}[ht]
{\centering \resizebox*{8cm}{9.5cm}{\includegraphics{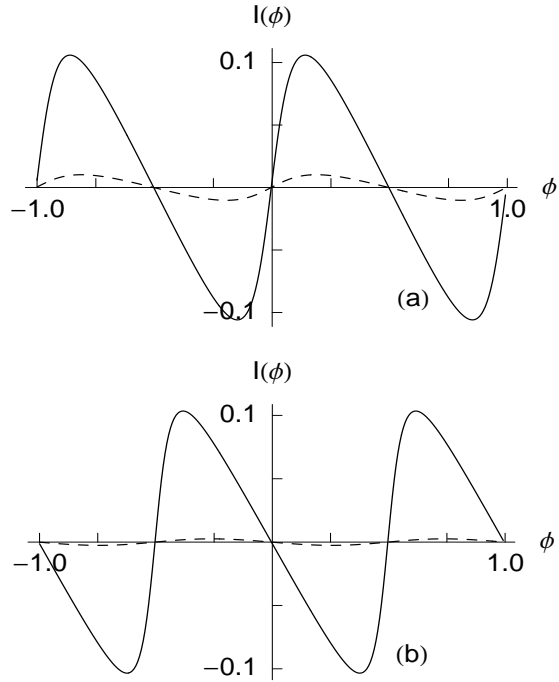}}\par}
\caption{Persistent current as a function of $\phi$ for the disordered rings 
with $N=100$, $\alpha=0.9$, $W=1$, and, a) $N_e=20$ and b) $N_e=15$. The
dotted and solid lines are respectively for the rings with NNH and 
SNH integrals.}
\label{current2}
\end{figure}

The results for the larger disordered rings are given in Fig.~\ref{current2}
where we take $N=100$, $\alpha=0.9$ and $W=1$. The persistent currents 
corresponding to the cases with NNH and SNH integrals are respectively 
represented by the dotted and solid lines. Here results are presented for 
some typical disordered configurations of the ring and in fact, we observe 
that the qualitative behavior of the persistent currents do not depend on 
the specific realization 
of the disordered configurations. This figure shows that the persistent 
currents for
the disordered rings are always periodic in $\phi$ with $\phi_0$ flux 
periodicity. In the presence of disorder, we see from Fig.~\ref{current2} 
that the persistent current always becomes a continuous function of magnetic 
flux $\phi$, and this behavior can be understood as follows 
(see Ref.~\cite{san2}). The sharp
transitions at the points $\phi=0$ or $\pm n\phi_0$ with
even $N_e$ and at $\phi=\pm n\phi_0/2$ with odd $N_e$, 
for the perfect rings (see Fig.~\ref{current1}) appear due to the 
\begin{figure}[ht]
{\centering \resizebox*{8.5cm}{9.5cm}{\includegraphics{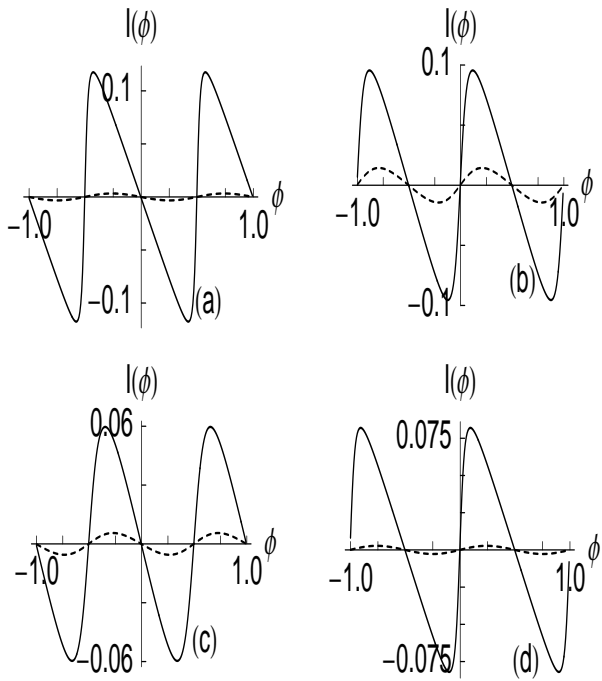}}\par}
\caption{Persistent current as a function of $\phi$ for different disordered  
($W=1$) rings with higher electron concentrations. Here we choose $\alpha=1.1$.
a) $N=125$, $N_e=45$; b) $N=125$, $N_e=40$; c) $N=150$, $N_e=55$ and 
d) $N=150$, $N_e=60$.} 
\label{oddeven}
\end{figure}  
degeneracy of the ground state energy at these points. 
Now as the impurities are introduced, all the
degeneracies get lifted and current exhibits a continuous variation with
respect to $\phi$. At these degenerate points, the ground state energy 
passes through an extrema which in turn gives zero persistent current 
as shown in Fig.~\ref{current2}. 
It is clear from Fig.~\ref{current2} that the higher order hopping
integral play an important role to enhance the amplitude of persistent
current in the disordered rings. From Fig.~\ref{current2}(a) and
Fig.~\ref{current2}(b) we see that the currents in the disordered 
rings with only NNH integrals (the dotted lines) are vanishingly small 
compared to those as observed in the impurity free rings with NNH integrals 
(the dotted curves in Fig.~\ref{current1}(a) and Fig.~\ref{current1}(b)). 
On the other hand, Fig.~\ref{current2} shows that the persistent currents
in the disordered rings with higher order hopping integral are of the 
same order of magnitude as those for the ordered rings. 

 In Fig.~\ref{oddeven} we give persistent currents for the disordered rings
with higher electron concentrations and study the cases with $N=$ even or
odd and $N_e=$ even or odd. The dotted and solid curves respectively
corresponds to the NNH and SNH models. It is observed that the evenness
or the oddness of $N$ and $N_e$ do not play any important role on persistent
current but we will see that the diamagnetic or paramagnetic sign of persistent
current crucially depends on the evenness or oddness of $N_e$.

Physically, the higher order hopping integrals
try to delocalize the energy eigenstates and thereby favor phase
coherence of the electrons even in the presence of disorder, and thus
prevents the reduction of persistent current due to disorder. While 
in the disordered rings with only NNH integrals, the enormous reduction of
current amplitudes are basically due to localization of the energy eigenstates.
When we add higher order hopping integrals, it is most likely that the 
localization length increases and may become comparable to the length of the
ring, and, we get enhancement of persistent current. 
\section{Mesoscopic Cylinder}
\begin{figure}[ht]
{\centering \resizebox*{5.75cm}{3.75cm}{\includegraphics{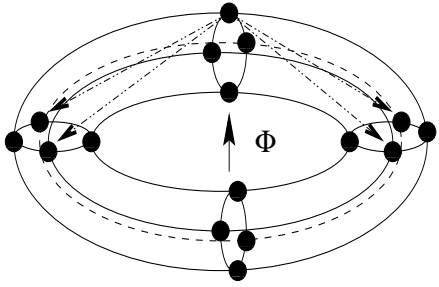}}\par}
\caption{A normal metal mesoscopic cylinder threaded by a magnetic flux $\phi$.
The filled circles correspond the positions of the lattice site.}
\label{cylinder}
\end{figure}

 This section investigates the behavior of persistent currents as a function
of magnetic flux $\phi$ both in perfect and dirty multi-channel mesoscopic 
cylinders described respectively by NNH and SNH (diagonal hopping shown by
the arrows in Fig.~\ref{cylinder}) integrals. The main motivation for 
the study of the characteristic behaviors of persistent current in these
mesoscopic cylinders is that, due to the existence of multi-channels, 
there is a possibility of diffusion of the electrons in presence of impurity
and the enhancement of persistent current in diffusive systems can be clearly
verified. 

The tight-binding Hamiltonian of such a multi-channel mesoscopic cylinder 
threaded by a magnetic flux $\phi$ with $N$ and $M$ number of sites 
respectively along the longitudinal and transverse direction can be written 
in the following form
\begin{eqnarray}
H &=& \sum_l \epsilon_l c_l^{\dagger}c_l + \sum_t \epsilon_t c_t^{\dagger}c_t
+ \sum_{<t,t^{\prime}>}v_{tt^{\prime}}c_t^{\dagger}c_{t^{\prime}} \nonumber \\
& & + \sum_{<l,l^{\prime}>}\left[v_{ll^{\prime}}e^{i\theta_{ll^{\prime}}}
c_l^{\dagger} c_{l^{\prime}} + h.c. \right] \nonumber \\
& & + \sum_{<d,d^{\prime}>}\left[v_{dd^{\prime}}e^{i\theta_{dd^{\prime}}}
c_d^{\dagger}c_{d^{\prime}} + h.c. \right]
\label{cylhamil}
\end{eqnarray}
where, $\epsilon_l$ and $\epsilon_t$ are the site potential energies along
the longitudinal and transverse direction respectively. $v_{tt^{\prime}}$
is the transverse hopping strength, while, $v_{ll^{\prime}}$ and
$v_{dd^{\prime}}$ respectively corresponds the hopping strength along the
longitudinal and diagonal direction. The phase factors $\theta_{ll^{\prime}}$
and $\theta_{dd^{\prime}}$ are identical with $\left(2\pi \phi/N\right)$ and
the diagonal hopping strength $v_{dd^{\prime}}=v\exp(-\alpha)$, $v$ is the 
NNH strength along the longitudinal direction.

Here we focus the behavior of persistent current for perfect and dirty
cylindrical systems considering both NNH and SNH integrals. For a perfect
cylinder, taking $\epsilon_l=0$ for all $l$ and $\epsilon_t=0$ for all $t$, 
the energy eigenvalue of $n$th eigenstate is expressed in the form
\begin{eqnarray}
E_n(\phi) &=& 2~v\cos\left[\frac{2\pi}{N}(n+\phi)\right] \nonumber \\
& & + 4~v\exp(-\alpha)\cos\left[\frac{2\pi}{N}(n+\phi)\right]\times 
\nonumber \\ 
& & \cos\left[\frac{2\pi m}{M}\right] + 2~v\cos\left[\frac{2\pi m}{M}\right]
\label{engequ2}
\end{eqnarray}
and the corresponding persistent current carried by this eigenstate is given by
\begin{eqnarray}
I_n(\phi) &=& \left(\frac{4\pi v}{N}\right)\sin\left[\frac{2\pi}{N}(n+\phi)
\right] \nonumber \\
& & +\left(\frac{8\pi v}{N}\right)\exp(-\alpha)\sin\left[\frac{2\pi}{N}(n+\phi)
\right] \times \nonumber \\
& & \cos\left[\frac{2\pi m}{M}\right]
\label{currequ2}
\end{eqnarray}
where $n$ and $m$ are two integers respectively bounded within the
range : $-\lfloor N/2 \rfloor \leq n < \lfloor N/2 \rfloor$ and
$-\lfloor M/2 \rfloor \leq m < \lfloor M/2 \rfloor$.

\begin{figure}[ht]
{\centering \resizebox*{8cm}{9.5cm}{\includegraphics{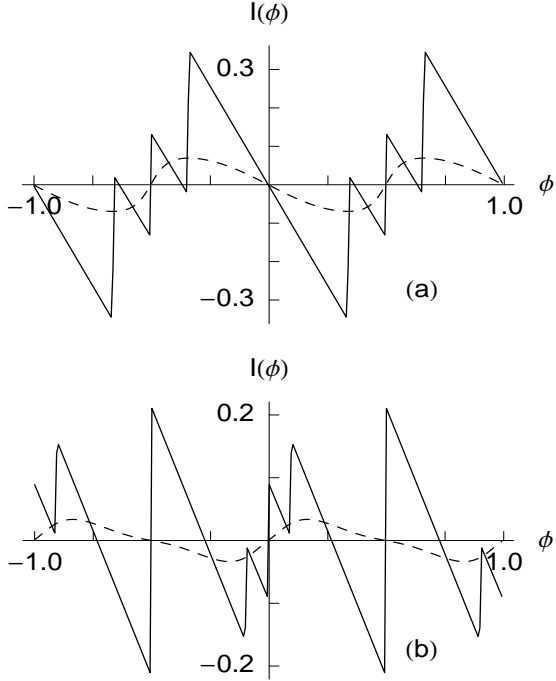}}\par}
\caption{Persistent current as a function of $\phi$ for multi-channel
mesoscopic cylinders described by only NNH integral with $N=50$, $M=4$,
and, a) $N_e=45$ and b) $N_e=40$. The solid and dotted lines 
are respectively for perfect ($W=0$) and dirty ($W=1$) cylinders.}
\label{cylnnhcurr}
\end{figure}

Now try to explain the behavior of persistent current in multi-channel
cylindrical systems described by NNH integral only. As representative
examples we plot the results of persistent current in these systems
in Fig.~\ref{cylnnhcurr}. Here we consider the ring size $N=50$ along the
longitudinal direction and $M=4$ along the transverse direction. The
results shown in Fig.~\ref{cylnnhcurr}(a) and Fig.~\ref{cylnnhcurr}(b)
are respectively for the cylinders with $N_e=45$ and $N_e=40$ where the
solid lines correspond the variation of persistent current in absence
of any impurity ($W=0$) and the dotted lines correspond those results
for the dirty systems with disorder strength $W=1$. Now in these 
multi-channel perfect systems current shows several kink-like structures
(see solid curves in Fig.~\ref{cylnnhcurr}(a) and Fig.~\ref{cylnnhcurr}(b)) 
at different values of $\phi$, depending on $N_e$, compared to the
results for one-channel perfect rings (see the curves in 
Fig.~\ref{current1}(a) and Fig.~\ref{current1}(b)). This is due to the
fact that in multi-channel systems several additional overlaps of the 
energy levels take place compared to the one-channel systems. But the
current always gets $\phi_0$ flux-quantum periodicity. As the impurities
are switched on all the degeneracis go out and current gets a continuous
variation as shown by the dotted curves in Fig.~\ref{cylnnhcurr}(a) and 
Fig.~\ref{cylnnhcurr}(b). For these cylindrical systems described by
NNH integral only it is observed that in presence of disorder current
amplitude gets reduced by an order of magnitude compared to the current
amplitude in perfect systems.

\begin{figure}[ht]
{\centering \resizebox*{8cm}{9.5cm}{\includegraphics{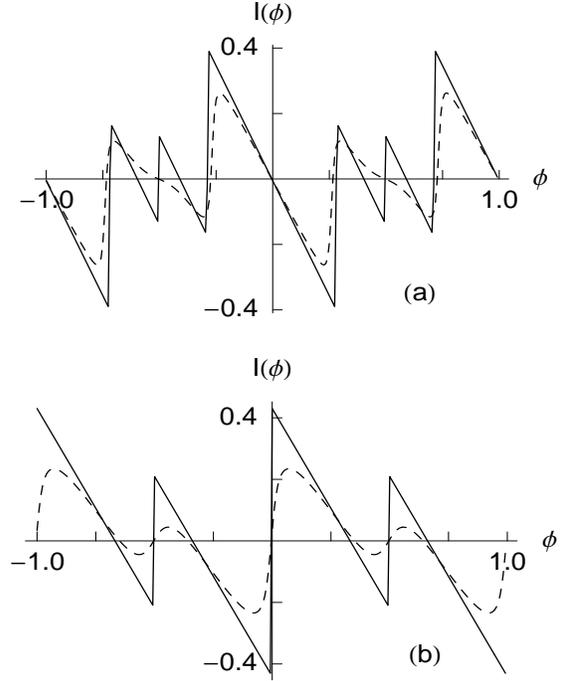}}\par}
\caption{Persistent current as a function of $\phi$ for multi-channel 
mesoscopic cylinders described by both NNH and SNH ($\alpha=1.0$) integrals 
with $N=50$, $M=4$, and, a) $N_e=45$ and b) $N_e=40$. The solid and dotted 
curves are respectively for perfect ($W=0$) and dirty ($W=1$) cylinders.}
\label{cylsnhcurr}
\end{figure}

Now we focus our attention on the behavior of persistent current for the
multi-channel cylindrical systems described with both NNH and SNH integrals.
In Fig.~\ref{cylsnhcurr}(a) and Fig.~\ref{cylsnhcurr}(b) we display the
variation of persistent currents as a function of $\phi$ for the multi-channel
mesoscopic cylinders in presence of SNH ($\alpha=1.0$) integral in addition 
to the NNH integral taking the same system size ($M=50$ and $N=4$) as in
the systems described with NNH integral only. The results shown in 
Fig.~\ref{cylsnhcurr}(a) and Fig.~\ref{cylsnhcurr}(b) are respectively
for the systems with $N_e=45$ and $N_e=40$ where the solid curves present
the results of perfect ($W=0$) cylinders and the dotted curves give
the results of dirty ($W=1$) cylinders.
From the curves shown in Fig.~\ref{cylsnhcurr}(a) and Fig.~\ref{cylsnhcurr}(b)
we can emphasize that current amplitudes in dirty systems (see dotted curves)
are comparable to that of perfect systems (see solid curves). This is due to
the fact that higher order hopping integrals try to delocalize the energy
eigenstates and thus current amplitude increases, even an order of
magnitude, in comparison with the current amplitude in dirty cylinders
described with NNH integral only.

 Thus our results for both one-channel mesoscopic rings and multi-channel 
mesoscopic cylinders predict that higher order hopping integral has an 
important role for the enhancement of current amplitude in presence of 
impurity.
\section{Low-field Magnetic Susceptibility}
 In this section we address the behavior of low-field magnetic 
susceptibility for both the ordered and disordered one-channel mesoscopic 
rings and multi-channel cylinders described by the Hamiltonians with NNH 
and SNH integrals. 
\begin{figure}[ht]
{\centering \resizebox*{8cm}{6cm}{\includegraphics{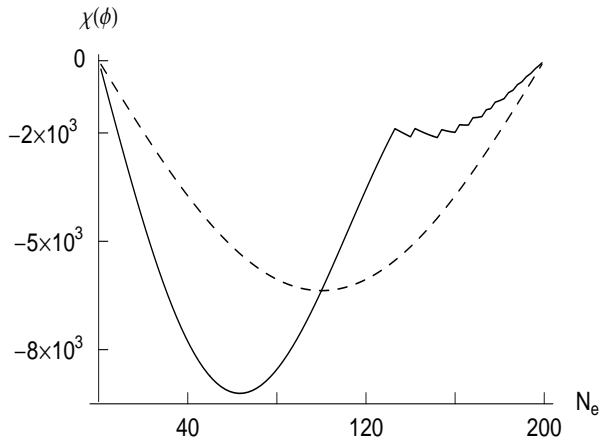}}\par}
\caption{$\chi$ versus $N_e$ curves near zero-field for ordered rings with 
$N=200$ and $\alpha=1.1$. The dotted and solid lines are respectively for 
the rings with NNH and SNH integrals.}
\label{suscep}
\end{figure}  
This quantity can be calculated from the first order derivative of persistent 
current and the general expression of magnetic susceptibility is of the form
\begin{equation}
\chi(\phi)=\frac{N^3}{16 \pi^2}\left(\frac{\partial I(\phi)}{\partial \phi}
\right)
\end{equation}

Calculating magnetic susceptibility we can  precisely predict the 
diamagnetic and paramagnetic signs of the persistent currents in such 
systems~\cite{levy,chand,yu,jari,deb}.
Our calculations for strictly one-channel rings show that the sign of the 
persistent current is not a random quantity, rather it is independent of 
the specific realizations of disorder, while, the calculations for 
multi-channel cylinders emphasize that the sign of the currents cannot be 
predicted precisely since it strongly depends on the total number of 
electrons, $N_e$, and also on the specific realization of disordered 
configurations of the system.

 Let us first try to describe the sign of the low-field currents in strictly
one-channel mesoscopic rings. In a perfect ring, the magnetic susceptibility 
\begin{figure}
{\centering \resizebox*{8cm}{10cm}{\includegraphics{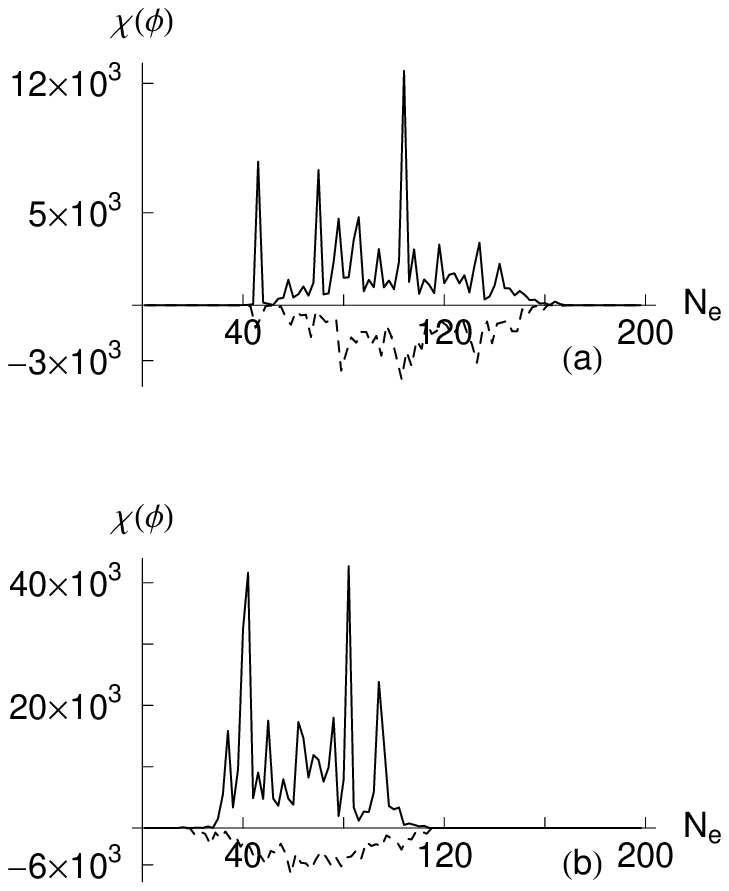}}\par}
\caption{$\chi$ versus $N_e$ curves near zero-field for disordered rings 
with $N=200$, $\alpha=1.1$ and $W=1$. The results corresponding to
the Hamiltonians with NNH and SNH integrals are
respectively presented in a) and b). The solid and dotted lines
are respectively for the rings with even and odd $N_e$.} 
\label{susdis}
\end{figure}  
associated with the current $I_n(\phi)$ carried by $n$th eigenstate can be 
expressed as
\begin{eqnarray}
\chi_n(\phi) &=& \frac{N}{4}\sum_{p=1}^{p_0} 2~v~p^2 \exp\left[\alpha 
(1-p)\right] \times \nonumber \\
& & \cos\left[\frac{2\pi p}{N}\left(n+\phi\right)\right]
\label{susequ1}
\end{eqnarray}
At zero
temperature the total magnetic susceptibility will be $\chi(\phi)=\sum_n
\chi_n(\phi)$, where the summation over the quantum number $n$ lies in the 
range $-\lfloor N_e/2\rfloor \leq n < \lfloor N_e/2 \rfloor$. 
In Fig.~\ref{suscep}, we display the variation of low-field magnetic 
susceptibility of perfect rings with the number of electrons $N_e$ 
in the rings. The dotted and solid curves respectively correspond  the 
rings described by the Hamiltonians with NNH and SNH integrals.   
These two curves indicate that in the limit $\phi\rightarrow 0$, persistent 
current exhibits a diamagnetic sign irrespective of the total number of 
electrons, $N_e$, in the rings. This diamagnetic sign of the currents follows 
from the slope of the curves at the zero field limit ($\phi\rightarrow 0$) 
presented in Fig.~\ref{current1}. So we conclude that at low magnetic fields 
($\phi \rightarrow 0$), there will be only diamagnetic persistent currents 
in perfect rings.

 Now we investigate the behavior of low-field magnetic susceptibility
for the disordered rings. In Fig.~\ref{susdis} we plot the 
low-field magnetic susceptibility as a function of $N_e$ for the rings
taking the ring size $N=200$ and disorder strength $W=1$. The results for 
the models with NNH 
and SNH integrals are displayed respectively in Fig.~\ref{susdis}(a),
and Fig.~\ref{susdis}(b) considering $\alpha=0.9$. The solid and dotted lines 
in these figures are
respectively for the rings with even and odd number of electrons, $N_e$. The
curves in Fig.~\ref{susdis} correspond to some typical disordered 
configurations of the rings. The most interesting finding is that
the persistent currents in the disordered rings always show diamagnetic 
sign for odd $N_e$ and paramagnetic sign for even $N_e$. At low fields
the $I-\phi$ curves for the perfect rings have a discontinuity when $N_e$ 
is even whereas for odd $N_e$ there is no discontinuity 
(see Fig.~\ref{current1}), but in both the cases persistent currents have
diamagnetic sign. As disorder removes the discontinuity of the $I-\phi$
curves, the slopes of the $I-\phi$ curves near zero field become positive
for even $N_e$ while the slopes remain negative for odd $N_e$ 
(see Fig.~\ref{current2}).
This has a very general consequence that irrespective of the disordered 
configurations, at low fields we always get diamagnetic persistent
current when $N_e$ is odd and paramagnetic current when $N_e$ is even.

 Finally, let us consider the behavior of the sign of persistent currents 
for the mesoscopic multi-channel cylinders in the limit $\phi \rightarrow 0$.
In absence of any impurity i.e., for perfect cylinders the magnetic
susceptibility associated with the current $I_n(\phi)$ carried by $n$th
energy eigenstate considering both NNH and SNH (diagonal hopping which
is shown by the arrows in Fig.~\ref{cylinder}) integrals is written in 
the form
\begin{eqnarray}
\chi_n(\phi) &=& \frac{Nv}{2}\left\{\cos\left[\frac{2\pi}{N}(n+\phi)\right] 
+2~\exp(-\alpha)\times \right. \nonumber \\
& & \left.\cos\left[\frac{2\pi}{N}(n+\phi)\right]\cos\left[\frac{2\pi m}{M}
\right]\right\} 
\label{susequ2}
\end{eqnarray}
In these cylindrical systems 
the sign of the low-field currents cannot be predicted exactly, even in
absence of any impurity, since the sign of the currents strongly depends 
on the total number of electrons, $N_e$, and for dirty systems it also
strongly depends on the specific realization of disordered configurations.
\section{Magnitude of Persistent Current Amplitude with System Size $N$}
\begin{figure}
{\centering \resizebox*{8cm}{9.5cm}{\includegraphics{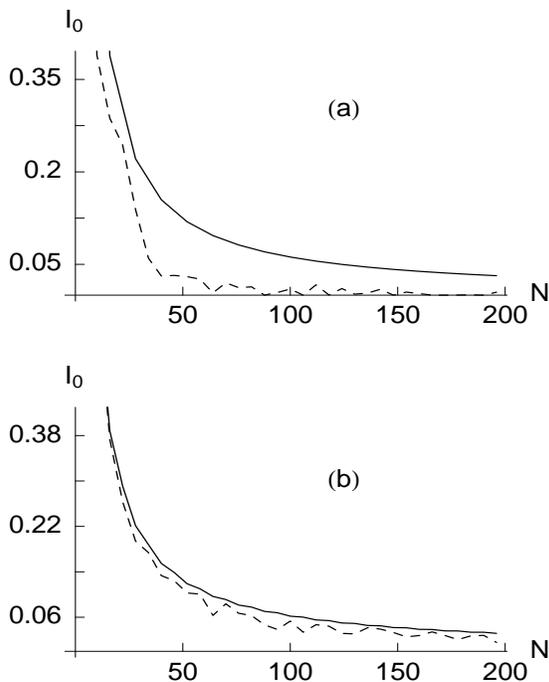}}\par}
\caption{Current amplitude $I_0$ as a function of system size $N$ in
one-channel mesoscopic rings keeping the ratio $N_e/N$ as a constant by 
the relation $N=2N_e$, where (a) rings described with NNH integral only
and (b) rings described with both NNH and SNH ($\alpha=0.9$) integrals.
The dotted and solid lines respectively corresponds the results for 
perfect and dirty ($W=1$) systems.}
\label{converge1}
\end{figure}

 In this section we show that the higher order hopping integrals play an 
important role
to enhance persistent current in the disordered mesoscopic rings and
cylinders. For this purpose we study the behavior of persistent current with 
system size $N$ in these systems for constant electron density, i.e., for
constant $N_e/N$ ratio. We have calculated the current amplitude $I_0$ 
at some typical field, say at $\phi=0.25$, and the $I_0$ versus $N$ curves
are shown in Fig.~\ref{converge1}. The results for the rings described 
with NNH integral only are plotted in Fig.~\ref{converge1}(a), while those 
for the rings described by NNH and SNH integrals are shown in 
Fig.~\ref{converge1}(b) keeping the ratio $N/N_e=2$ in both the cases. 
The dotted and solid lines correspond to the rings in absence of any impurity
($W=0$) and in presence of impurity with strength $W=1$ respectively.
If we compare Fig.~\ref{converge1}(a) and Fig.~\ref{converge1}(b), we see
that the currents in the disordered rings are orders of magnitude less than 
those for the perfect rings at the mesoscopic length scale when we use the
model with only NNH integrals (see dotted curve in Fig.~\ref{converge1}(a)). 
Quite interestingly we 
observe that when the SNH integrals are switched on in addition
to the NNH integrals in the disordered rings, the current amplitudes have 
some finite non-zero values, comparable to the perfect ring results,
even if $N$ is in the mesoscopic regime and this is evident from the
dotted curve of  Fig.~\ref{converge1}(b).

In multi-channel mesoscopic cylinders, keeping $N_e/N$ ratio as a constant,
we also get the similar kind of behavior for the amplitude variation.

In the nearest-neighbor tight-binding model, disorder tries to localize
the electrons and the persistent current becomes almost zero at the
mesoscopic length scale. On the other hand, in the presence of higher 
order hopping integrals the electron eigenstates are not localized within
the mesoscopic length scale, and we get enhanced persistent current as
the electronic phase coherence is preserve over the sample size.
Accordingly, we can emphasize that both for one-channel mesoscopic rings 
and multi-channel mesoscopic cylinders the higher order hopping integrals
have an important role for the enhancement of persistent current amplitude.
\section{Conclusion}
 In conclusion, we have studied in details the characteristic behavior of
persistent current and low-filed magnetic susceptibility in one-channel
mesoscopic rings and multi-channel mesoscopic cylinders within the 
tight-binding framework in presence of higher order hopping integrals. 
We have shown that the addition of higher order hopping integrals in the 
nearest-neighbor tight-binding Hamiltonian gives order of magnitude 
enhancement of persistent current in the disordered mesoscopic rings and
cylinders. In this paper we have also calculated low field magnetic 
susceptibility of these systems as a function of $N_e$, and our exact 
calculations for one-channel rings show that the sign of the current 
is independent of the realization of disorder, it can be diamagnetic or 
paramagnetic depending on whether $N_e$ is odd or even, while the 
calculations for the multi-channel mesoscopic cylinders indicate that 
the sign of the low-field currents cannot be predicted, even in absence of
any impurity, since it strongly depends on $N_e$ and for dirty cylinders
it also depends on the specific realization of disordered configurations. 
From the variation of current amplitude with system size $N$ for constant 
electron density, we see that enhancement of persistent current due to 
higher order hopping integrals will be appreciable only in the mesoscopic 
scale.

\end{document}